\documentclass[12pt,preprint]{aastex}

\usepackage{times}
\usepackage{lscape}

\newcommand{\ovi}{O$\;${\small\rm VI}\relax}

\newcommand{\sii}{S$\;${\small\rm II}\relax}

\newcommand{\sone}{S$\;${\small\rm I}\relax}
\newcommand{\sthree}{S$\;${\small\rm III}\relax}
\newcommand{\sfour}{S$\;${\small\rm IV}\relax}
\newcommand{\ssix}{S$\;${\small\rm VI}\relax}

\newcommand{\othree}{O$\;${\small\rm III}\relax}

\newcommand{\lyalpha}{Lyman-$\alpha$}

\newcommand{\spp}{S$^{+2}$}

\newcommand{\electron}{$e^-$}

\newcommand{\htwo}{H$_2$}
\newcommand{\HI}{H$\;${\small\rm I}\relax}
\newcommand{\HII}{H$\;${\small\rm II}\relax}
\newcommand{\lya}{Lyman-$\alpha$}

\newcommand{\halpha}{H$\alpha$}

\newcommand{\Nelectron}{\ensuremath{N(e^-)}}

\newcommand{\kms}{km~s$^{-1}$\relax}
\newcommand{\vlsr}{$v_{\rm LSR}$\relax}

\newcommand{\percc}{cm$^{-3}$\relax}
\newcommand{\column}{cm$^{-2}$}

\newcommand{\nav}{$N_a(v)$}

\newcommand{\e}[1]{10^{#1}}

\newcommand{\z}{$z$}

\newcommand{\wave}[1]{$\lambda$#1\relax}
\newcommand{\twowave}[1]{$\lambda \lambda$#1\relax}

\newcommand{\fuse}{{\em FUSE}}

\newcommand{\hst}{{\em HST}}

\newcommand{\subwim}{\ensuremath{_{WIM}}}
\newcommand{\subhim}{\ensuremath{_{HIM}}}

\newcommand{\vz}{M~3--vZ~1128}
\newcommand{\zng}{M~5--ZNG~1}
\newcommand{\HSS}{Howk et al. (2006)}
\newcommand{\dm}{\ensuremath{{\rm DM}}}

\begin{document}


\title{Ionized Gas in the First 10 Kiloparsecs of the Interstellar
  Galactic Halo: Metal Ion Fractions\altaffilmark{1}}


\author{J. Christopher Howk\altaffilmark{2}, 
  S. Michelle Consiglio\altaffilmark{2,3}}

\altaffiltext{1}{Based on observations with the NASA/ESA {\em Hubble
    Space Telescope} obtained at the Space Telescope Science
  Institute, which is operated by the Association of Universities for
  Research in Astronomy, Inc., under NASA contract NAS5-26555.  These
  observations are associated with programs GO9150 and GO9410.  Also
  based on observations made with the NASA-CNES-CSA {\em Far
    Ultraviolet Spectroscopic Explorer}. FUSE was operated for NASA by
  the Johns Hopkins University under NASA contract NAS5-32985.}

\altaffiltext{2} {Department of Physics, University of Notre Dame,
  Notre Dame, IN, 46556; jhowk@nd.edu} \altaffiltext{3} {Current
  Address: Department of Physics and Astronomy, University of
  California, Los Angeles, Los Angeles, CA 90095,
  smconsiglio@ucla.edu}

\begin{abstract}

  We present direct measures of the ionization fractions of several
  sulfur ions in the Galactic warm ionized medium (WIM).  We obtained
  high resolution ultraviolet absorption line spectroscopy of
  post-asymptotic giant branch stars in the the globular clusters
  Messier~3 [$(l,b)=(42\fdg2, +78\fdg7); \ d=10.2 \ {\rm kpc}, z=10.0
    \ {\rm kpc}$] and Messier~5 [$(l,b) = (3\fdg9, +46\fdg8); d=7.5
    \ {\rm kpc}, z = +5.3 \ {\rm kpc}$] with the {\em Hubble Space
    Telescope} and {\em Far Ultraviolet Spectroscopic Explorer} to
  measure, or place limits on, the column densities of \sone, \sii,
  \sthree, \sfour, \ssix, and \HI. These clusters also house
  millisecond pulsars, whose dispersion measures give an electron
  column density from which we infer the \HII\ column in these
  directions.  We find fractions of S$^{+2}$ in the WIM for the M~3
  and M~5 sight lines $x(\mbox{\spp}) \equiv N(\mbox{\spp})/N({\rm S})
  = 0.33\pm0.07$ and $0.47\pm0.09$, respectively, with variations
  perhaps related to location.  With negligible quantities of the
  higher ionization states, we conclude S$^{+}$ and \spp\ account for
  all of the S in the WIM.  We extend the methodology to study the ion
  fractions in the warm and hot ionized gas of the Milky Way,
  including the high ions Si$^{+3}$, C$^{+3}$, N$^{+4}$, and O$^{+5}$.
  The vast majority of the Galactic ionized gas is warm ($T\sim10^4$
  K) and photoionized (the WIM) or very hot ($T>4\times10^5$ K) and
  collisionally ionized.  The common tracer of ionized gas beyond the
  Milky Way, O$^{+5}$, traces $<1\%$ of the total ionized gas mass of
  the Milky Way.

\end{abstract}


\section{Introduction}
\label{sec:intro}

Since its discovery in the 1970's, the origin of the diffuse \halpha\
emission arising from the Galaxy has remained something of a mystery
(Haffner et al. 2009, Reynolds 1993).  This emission, in addition to
the free-free radio absorption measurements from the 1960s (see Hoyle
\& Ellis 1963), implies the existence of a diffuse distribution of
free electrons outside of normal \HII\ regions.  This warm ionized
medium (WIM), often referred to as the ``Reynolds Layer'' in the Milky
Way, dominates the mass of ionized gas in the Milky Way and other
galaxies.  The power required to keep this gas ionized can be met
comfortably only by the Lyman-continuum photon production by
early-type stars (Haffner et al. 2009, Reynolds 1993).  However, the
ionization of this material by OB stars is troublesome: the large
scale height of the material ($\sim1$ kpc or more; Haffner et
al. 1999, G\'{o}mez et al. 2001, Gaensler et al. 2008) requires that
the photons responsible for ionizing the WIM travel hundreds of
parsecs from their point of origin.  The large cross-section for
absorption of such photons by neutral hydrogen ($\sigma \sim
6.3\times10^{-18}$ cm$^2$) naively implies such photons could only
travel very small distances, of order $0.1$ pc assuming a typical
interstellar density ($\sim1$ \percc). Various groups have presented
models for the ionization of the WIM which rely on photons from hot
stars (Mathis 2000; Sembach et al. 2000; Domg\"{o}rgen \& Mathis
1994), photon emission from cooling hot gas (Slavin et al. 2000),
heating from magnetic reconnection (Hoffman et al. 2012), and even
photons from decaying neutrinos (Sciama 1998).

The hot star ionization models, and to a lesser extent the models
invoking cooling hot gas, must assume that the geometry of the
interstellar medium (ISM) allows for the propagation of ionizing
photons over the large distances required (Miller \& Cox 1993, Dove \&
Shull 1994, Dove et al. 2000, Wood \& Mathis 2004, Wood et al. 2010).
This includes the concept of density-bounded or ``leaky'' \HII\
regions (Haffner et al. 2009).  In particular, various groups have
noted the potential importance of the complicated geometries that may
exist in a turbulent, supernova-driven ISM (Miller \& Cox 1993, Wood
\& Mathis 2004), or in superbubbles, where a significant amount of gas
has been evacuated about OB associations as a result of correlated
supernova explosions (Wood et al. 2010, Dove \& Shull 1994, Dove et
al. 2000).  Coupled with ``leaky'' \HII\ regions, these geometric
considerations may go a long way toward explaining the propagation of
ionizing photons from their sources to the WIM gas.  However, there is
not yet a complete model of the ionization of the WIM, and the
predicted ionizing spectrum depends on the assumed sources and the degree
to which the photons may be processed through intervening H, thereby
modifying the source spectrum (Haffner et al. 2009).

Very sensitive observations of forbidden metal emission lines acquired
over the last decade with the Wisconsin H-alpha Mapper (WHAM; Reynolds
et al. 1998b; Tufte 1997, Haffner 1999, Madsen et al. 2006) have
provided data on the ionization and temperature structure of the WIM
(Reynolds et al. 1998a, Haffner et al. 1999, Reynolds et al. 2001,
Hausen et al. 2002, Madsen et al. 2006).  Compared with classical
\HII\ regions the WIM appears to be warmer (Madsen et al. 2006).
Metals and helium in the WIM are generally less highly ionized than in
\HII\ regions, as well.  In particular, the weakness of forbidden
[\ion{O}{3}] and \ion{He}{1} recombination emission implies relatively
low ionization fractions of O$^{+2}$ and He$^+$ in the WIM compared
with classical \HII\ regions (Reynolds \& Tufte 1995, Madsen et
al. 2006).  Haffner et al. (1999) demonstrated that the ratio of
forbidden metal line strengths [\ion{S}{2}]/[\ion{N}{2}] is sensitive
to the ionization fraction of S$^+$: $x({\rm S}^+) \equiv n({\rm
  S}^+)/n({\rm S})$.  The typical values implied for this ionization
fraction in the WIM are $x({\rm S}^+) \approx 0.3-0.7$ (Haffner et
al. 1999, Madsen et al. 2006).  This is higher than seen in O star
\HII\ regions where \spp\ is more abundant than S$^+$.  The emission
line constraints on WIM ionization imply a lower ionization parameter
for the WIM compared to classical \HII\ regions and a distinct
spectrum, e.g., due to spectral processing of O star radiation
escaping from \HII\ regions and/or contributions from other sources
such as cooling radiation (e.g., Slavin et al. 2000) or hot, evolved
stars such as white dwarfs (Bregman \& Harrington 1986, Haffner et
al. 2009, Flores-Fajardo et al. 2011).

While emission line diagnostics have provided some quantitative
measures of the WIM ionization, their temperature dependence adds a
layer of complexity to understanding the ionization of the medium.
Furthermore, because the excitation of these lines relies on
collisions with warm electrons, their intensities are weighted by
$\approx n_e^2$, where $n_e$ is the electron density, implying they
may be strongly weighted toward higher density regions.  In this paper
we present a complementary approach to probing the ionization state of
the WIM.  Following Howk et al. (2006) we use ultraviolet (UV)
absorption line measures of the multiphase ISM along sight lines to
globular clusters that also contain pulsars.  The UV observations of
post-asymptotic giant branch (PAGB) stars in the background clusters
from the {\em Hubble Space Telescope} (\hst) and {\em Far Ultraviolet
  Spectroscopic Explorer} (\fuse) provide measures of metal ion column
densities in these directions, while radio observations of pulsar
dispersion measures (Hessels et al. 2007) provide determinations of
the electron column densities.  Taken together we can provide
estimates of the ionization fractions of S$^0$, S$^{+}$, S$^{+2}$,
S$^{+3}$, and S$^{+5}$ in the WIM (as well as the total gas phase
abundances; see Howk et al. 2006).  We apply this technique to two
extended sight lines through the Galactic WIM, one probing the first
10 kpc above the disk in the very local region about the sun (Howk et
al. 2003, 2006) and the other probing the first 5 kpc of the disk, but
some 5 kpc projected radial distance toward Galactic Center (Zech et
al. 2008).

The structure of our paper is as follows.  In \S \ref{sec:method} we
discuss the methodology underlying our technique, which builds upon
the discussion of elemental abundance determinations in \HSS.  In \S
\ref{sec:observations} we discuss the UV absorption line observations,
their reduction, and our assessment of the hydrogen and metal column
densities along the two sight lines in this work.  In \S
\ref{sec:analysis} we present the results of our analysis of these two
sight lines and compare these with previous constraints on WIM
physical conditions, while in \S \ref{sec:discussion} we discuss the
implications of our results and compare them with theoretical models.
We give a summary of our results in \S \ref{sec:summary}.

\section{Methodology}
\label{sec:method}

To provide a direct quantitative measure of the ionization fractions
of metal ions in the WIM, we rely on unique sight lines that provide
measures of both metal ions and neutral hydrogen from UV absorption
lines as well as ionized hydrogen from pulsar dispersion measures
(DMs).  A comparison of the metal ions with the \HII\ column density
derived from the pulsar dispersion measures gives the ionization
fractions if the abundance of the metal is known.

We write the abundance of a metal $X$ with respect to H: 
\begin{equation}
\label{eqn:simpleabundance}
        A(X) \equiv \frac{\sum\limits_j N(X^j)}
                { N(\mbox{\HI}) + N(\mbox{\HII}) + 2N({\rm H}_2)},
\end{equation}
where $N(X^j)$ is the column density of the $j$th ionization stage of
$X$ and the sum is nominally over all ionization states.  Thus, the
numerator represents the total column density of the metal $X$, while
the denominator, with $N({\rm H}) \equiv N(\mbox{\HI}) +
N(\mbox{\HII}) + 2N({\rm H}_2)$, is the total hydrogen column density.
In most interstellar absorption line studies, $A(X)$ is estimated by
comparing the dominant ionization state of the element $X$ in the warm
neutral medium (WNM) with the column density of \HI.  Such studies
assume that the neglect of terms in the sums in both the numerator and
denominator of Equation \ref{eqn:simpleabundance} has a negligible
effect, which is not necessarily true (Sembach et al. 2000), or they
attempt to make ionization corrections on the basis of models (e.g.,
Howk et al. 1999, Howk et al. 2003).  Howk et al. (2006) showed that
all of the significant terms can be accounted for along sight lines to
globular clusters with UV bright stars and radio pulsars. Ultraviolet
absorption lines provide measurements of all of the important ionic
states of S and sometimes other metals in the WIM and WNM (\S
\ref{sec:observations}).  Those UV measurements combined with the
pulsar dispersion measures provide information on all of the states of
hydrogen.

Once the abundance is known, the comparison of the column of a metal
ion $X^j$ that arises in the WIM (i.e., with no contribution from the
the WNM) with the WIM \HII\ column density provides a measure of the
ionization fraction of $X^j$ in the WIM, $x(X^j) \equiv N(X^j)/N(X)$.
Thus, we can compare the ion S$^{+2}$ with the hydrogen reference for
the WIM, $N(\mbox{\HII})\subwim$:
\begin{eqnarray}
  \frac{x({\rm S}^{+2})}{x({\rm H}^+)} & = & 
  \frac{N(\mbox {\sthree})}{N({\rm S})\subwim} 
  \left[ \frac{N(\mbox{\HII})\subwim} 
    {N({\rm H})\subwim} \right]^{-1} \nonumber \\
  & = & \frac{N(\mbox {\sthree})}{N(\mbox {\HII})\subwim} A({\rm S})^{-1},
\label{eqn:himelectrons}
\end{eqnarray}
giving a measure of the ionization fraction of S$^{+2}$ to that of
H$^+$.  Here $N(\mbox{\sthree})$ does not require the WIM subscript as
it arises only within the WIM, with an energy for production that
precludes it arising in the WNM.  One may write the ionization
fractions for S$^{+3}$ or other WIM ions in a similar way.  We can
isolate $x({\rm S}^{+2})$ with a value of $x({\rm H}^+)=0.95\pm0.05$
based on observations of the weakness of [\ion{O}{1}] emission from
the Galactic WIM (Reynolds 1989, Reynolds et al. 1998a, Haffner et
al. 2009).

The determination of the \HII\ column in the WIM,
$N(\mbox{\HII})\subwim$, requires two corrections to the electron
column density, $N({\mbox \electron})$, provided by the DMs (Howk et
al. 2006). The first corrects for the contribution to $N({\mbox
  \electron})$ from hot ionized gas since we cannot measure the S ions
found in this hot gas (e.g., \ion{S}{7} and higher).  This correction
is discussed with the pulsar DMs below in \S \ref{sec:HIM}.  The other
is the correction for the \electron\ contributed to the DM from
ionized He.  We use the helium ionization correction factor, $\eta$,
discussed in detail in Howk et al. (2006) such that
$N(\mbox{\HII})=\eta \Nelectron$.  In this work we adopt $\eta_{WIM} =
0.98\pm0.01$, the ``minimum helium ionization'' case discussed in Howk
et al., which assumes the ionization fraction of He$^{+2}$ in the WIM
is minimal.  The choice of the minimum or maximum helium ionization
correction represents a (small) systematic uncertainty.  However, the
maximum helium ionization case ultimately produces an inconsistency:
we show below (even using the maximum case) that the ionization
fraction of He$^{+2}$ must in fact be very small (\S
\ref{sec:analysis},\S \ref{sec:discussion}).

\subsection{Correcting for the Hot ISM}
\label{sec:HIM}

We estimate HIM contribution to \Nelectron\ following the general
outline of Howk et al. (2006).  We assume the HIM electrons arise in
two components, one from T $\approx 10^5-10^6$ K gas traced by \ovi\,
the other from T$ > 10^6$ K gas traced by X-ray absorbing/emitting
gas.  Thus
\begin{equation}
\Nelectron\subhim  = 
    \eta\subhim^{-1} \left[ N(\mbox{\HII})_5 + N(\mbox{\HII})_6 \right],
\label{eqn:himelectrons1}
\end{equation}
where $N(\mbox{\HII})_5$ and $N(\mbox{\HII})_6$ are the \HII\ columns
associated with the $T\sim 10^{5-6}$ K and $>10^6$ K gas,
respectively.  The helium correction factor is $\eta\subhim$.  In this
coronal gas, we assume all of the helium is in the form of He$^{+2}$,
giving $\eta\subhim = 0.83$ (Howk et al. 2006).

We estimate the first term in brackets from Equation
\ref{eqn:himelectrons1} from the \ovi\ column density along the line
of sight (Wakker et al.  2003; Savage et al.  2003).  We write
\begin{equation}
  N(\mbox{\HII})_5 = \frac{N(\mbox{\ovi})}{A({\rm O}) x({\rm O}^{+5})},
\label{eqn:himelectrons2}
\end{equation}
where $N(\mbox{\ovi})$ is the measured \ovi\ column density derived
directly from \fuse\ data (see \S \ref{sec:metalcolumns}), $A({\rm O})
= (4.90\pm0.6)\times\e{-4}$ (Asplund et al. 2009) is the assumed
gas-phase oxygen abundance, and $x({\rm O}^{+5}) = 0.2$ (Sutherland \&
Dopita 1993, Savage et al. 2003) is the ionization fraction of
O$^{+5}$ in this phase of the gas (see discussion in Howk et
al. 2006).  This is the maximum value found in most models, and some
of the gas probed by \ovi\ could have an ionization fraction lower by
a factor of a few, depending on the conditions probed by the gas.
This lower-temperature component is the smallest contributor to the
whole, representing $<25\%$ of the total HIM column (in our
calculations and \S \ref{sec:analysis}).  We adopt a 50\% uncertainty
on the implied column given the large uncertainties involved.

The $T>10^6$ K gas, represented by the second term in brackets from
Equation \ref{eqn:himelectrons1}, has no direct probe along the
specific sight lines studied in this work.  Instead, we adopt a mean
Galactic distribution for this gas constructed to match X-ray
absorption measurements of \ion{O}{7}, \ion{O}{8}, and \ion{Ne}{9}
(Yao \& Wang 2005; Yao et al. 2009).  The absorption line measurements
are made toward both Galactic and extragalactic objects.  The refined
model presented in Yao et al. (2009) assumes a thick disk structure
described by a single exponential disk, $n_H(z) = n_H(0) \exp
(-|z|/h_z)$, with a mid-plane density $n_H(0) = 1.4\times10^{-3}$
\percc\ and scale height $h_z = 2.8$ kpc.  We integrate the $z$
distribution of hot gas densities to give
\begin{equation}
  N(\mbox{\HII})_6 = (\sin |b|)^{-1} \int_0^{z_*} n_H(z) dz
\label{eqn:himelectrons3}
\end{equation}
with the upper integration limit being the $z$-height of the star.
This estimate of the HIM column density has many simplifying
assumptions whose validities are difficult to assess.  As a result, we
adopt a 50\% uncertainty in our HIM column density estimates.

\section{Observations, Reductions, and Measurements}
\label{sec:observations}

In this work we apply the approach outlined above to study the
ionization fractions of S$^0$, S$^{+}$, S$^{+2}$, S$^{+3}$, and
S$^{+5}$ in the WIM for two directions through the Galactic WIM, those
toward the globular clusters M~3 [$(l,b)=(42\fdg2, +78\fdg7); \ d=10.2
  \ {\rm kpc}, z=10.0 \ {\rm kpc}$] and M~5 [$(l,b)=(3\fdg9,
  +46\fdg8); \ d=7.5 \ {\rm kpc}, z=+5.3 \ {\rm kpc}$].  In what
follows, we describe the UV data and analysis used to derive the metal
and neutral hydrogen column densities along these sight lines (\S 3.1
through \S 3.3).  We also describe the pulsar dispersion measures and
their uncertainties (\S 3.4).

\subsection{UV Observations}

In this work we make use of UV spectra from the Space Telescope
Imaging Spectrograph (STIS) on board \hst\ and from \fuse\ to study
the metal ion and \HI\ column densities toward M~3 and M~5.  The sight
line to M~3 is probed toward the PAGB star von Zeipel 1128 (vZ~1128);
the observations, reductions, and measurements for this sight line
have been discussed in Howk et al. (2003, 2006).  The ISM toward M~5
is probed along the sight line to the PAGB star ZNG~1; the
observations and their reductions have been described by Zech et
al. (2008).  The latter work specifically studied the high velocity
gas along the \zng\ direction.  We present the first measurements of
the low velocity gas along this sight line here.

The STIS\footnote{Proffitt et al. (2002) discuss the STIS instrument
  characteristics in detail.} observations used in this work all
employed the E140M grating to cover the spectral range $\sim$1150 to
1710 \AA\ at a resolution $R\equiv \lambda/\Delta \lambda \approx
45,800$.  This provides measures of the ISM absorption lines at an
equivalent velocity resolution $\Delta v \approx 6.5$ \kms\ (FWHM).
The \vz\ observations (program ID 9150; PI Howk) were obtained with
the $0\farcs2\times 0\farcs06$ aperture.  The typical signal-to-noise
ratios are in the range $\approx20$--40 per resolution element.  The
\zng\ observations (program ID 9410; PI Howk) were obtained with the
$0\farcs2\times 0\farcs2$ aperture.  The two datasets have slightly
different line spread function (LSF) shapes, notably differing in the
strength of the extended wings of the LSF, due to the different
apertures used.  The STIS \zng\ data have signal-to-noise ratios
$\ga25$ per resolution element in the regions of interest for this
work (Zech et al. 2008).

The \fuse\ observations used here cover the spectral range 905 to 1185
\AA\ at a resolution $R \approx 15,000$ giving a velocity resolution
$\Delta v \sim 20$ \kms\ (FWHM).  The data were all taken through the
LWRS $30\arcsec\ \times 30\arcsec$ apertures, and we have combined
observations acquired as part of several programs (Howk et al. 2003,
Zech et al. 2008).  These data have signal-to-noise ratios $\ga20$ per
resolution element in the regions of interest.  While the wavelength
calibration of the STIS data is excellent, the \fuse\ absolute (and to
some extent relative) wavelength calibrations are not as well
constrained.  As discussed in the earlier source papers (Zech et
al. 2008, Howk et al. 2006), we have bootstrapped the \fuse\
wavelength scale to match that of the STIS data using both stellar and
interstellar lines to determine the alignment.

\subsection{Metal Ion Column Densities}
\label{sec:metalcolumns}

We use the STIS and \fuse\ observations to derive column densities and
limits for several metal species toward \zng, with the relevant
results given in Table \ref{tab:columns}.  We focus on the ions of S,
since we have access to absorption from ions that probe the WIM
directly (\sthree\ and \sfour).  In addition, we give measurements of
the ``high ions'' \ion{Si}{4}, \ion{C}{4}, \ion{N}{5}, and \ovi\ along
both sight lines.  The \ovi\ is required for estimating the HIM \HII\
column density, but we will eventually use these high ions to
understand the ionization fractions of both the WIM and HIM in the
Galaxy (\S \ref{sec:analysis}).

To determine the metal ion columns we fit the stellar continuum in the
regions surrounding metal absorption lines using low order Legendre
polynomials.  We directly integrate the line absorption profiles and
apparent optical depths to determine the equivalent widths and
apparent column densities, $N_a$, following Sembach \& Savage (1992).
In all cases we adopt central wavelengths and oscillator strengths
from Morton (2003).  To calculate the limiting equivalent width and
apparent column density of \ion{S}{1}, we assume an intrinsic width of
$\sim30$ \kms\ (FWHM), the value derived from a single Gaussian
profile fit to the \sii\ 1250.584 \AA\ transition.  All limits given
in this work are $3\sigma$.

\begin{figure}
\epsscale{0.6}
\plotone{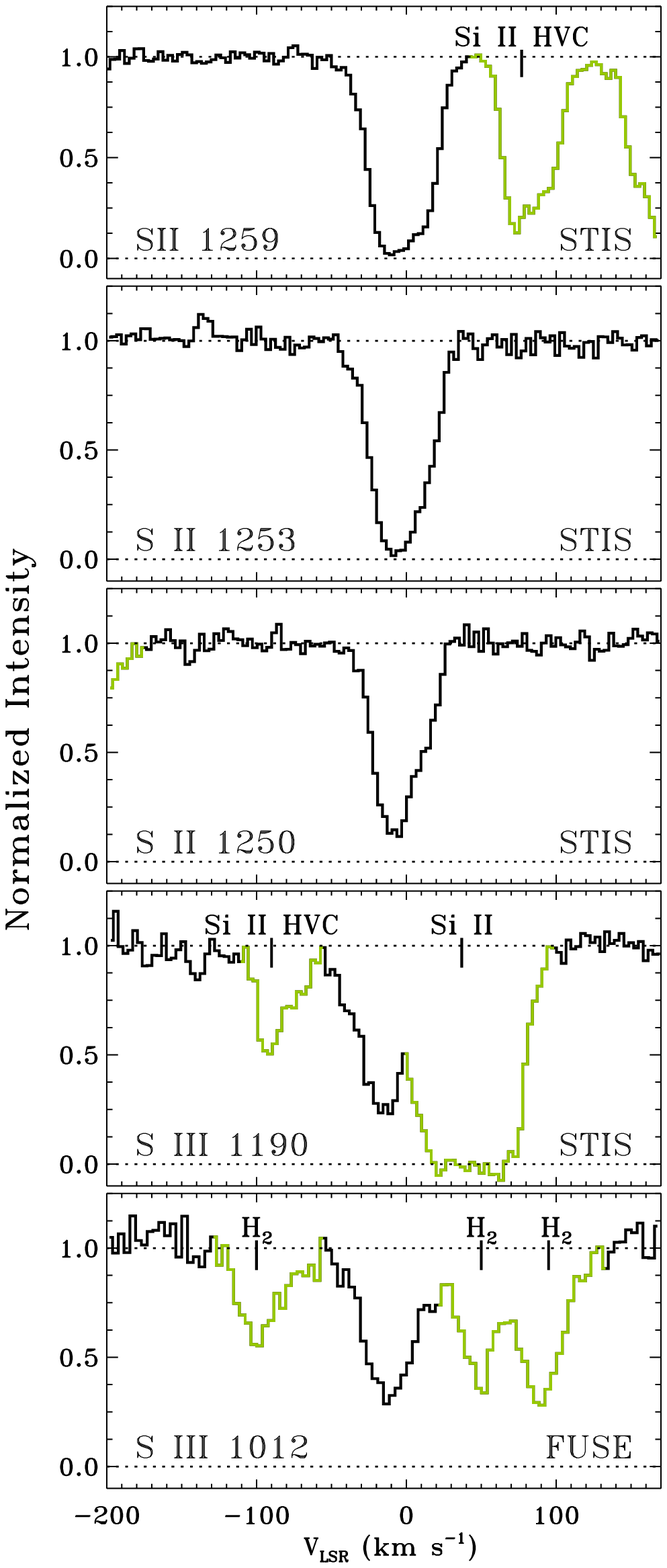}
\label{fig:spectra}
\caption{Absorption line profiles of \sii\ \twowave\ $1250.584$,
  $1253.811$, $1259.519$, and \sthree\ \wave\ $1190.208$ from STIS
  E140M observations as well as \sthree\ \wave\ $1012.495$ from
  \fuse\ observations of M5-ZNG1. The lighter colored portion of the
spectra are interloping lines. The \fuse\ data are shown with a binned
pixel size of $3.84$ \kms, giving $\sim 5$ pixels per resolution
element. The STIS data are shown with their native pixels, that is
$3.22$ \kms, or $\sim 2.3$ pixels per resolution element at these
wavelengths (Proffitt et al. 2002).}
\end{figure}

Deriving the column densities of \sii\ and \sthree\ toward
\zng\ required special care.  The absorption profiles for all of the
transitions from these ions are shown in Figure \ref{fig:spectra}.
The integrated apparent column densities derived from the weaker
\sii\ transitions at 1250.584 and 1253.811 \AA, which have $f$-values
that are different by a factor of two, differ by $\sim0.07$ dex, with
the weaker transition giving a higher $N_a$.  The strongest of the
\sii\ lines observed by STIS, at 1259.519 \AA\ and having an $f$-value
three times higher than the weakest, gives a significantly lower value
yet for $N_a$.  This progression of decreasing $N_a$ with increasing
$\lambda f$ suggests these lines contain unresolved saturated
structure (Savage \& Sembach 1991).

Figure 2 compares the apparent column density profiles as a function
of velocity, $N_a(v)$, for \sii\ $\lambda$1250, 1253, and
\sthree\ $\lambda$1190 (all from STIS).  The weakest \sii\ transition
(at 1250.584 \AA) is shown in each of the two panels as the thin black
histogram.  The discrepancies between the two \sii\ transitions in the
top panel are consistent with expectations in the classical case of
unresolved saturation.  Since there appears to be only moderate
saturation in the \sii, following Savage \& Sembach (1991) we correct
the column density for \sii\ $\lambda$1250 based on a comparison with
the 1253 transition and adopt an uncertainty appropriate for the
uncertainties in the method (the integrated uncertainties in $N_a(v)$
being minimal).  This gives our final adopted value of $\log
N(\mbox{\sii}) = 15.59\pm0.10$.  This amounts to an $+0.07$ dex
correction for saturation compared with the value derived from
\sii\ $\lambda$1250.

The \sthree\ transitions toward \zng\ are both contaminated to varying
degrees.  \sthree\ 1012 is contaminated by adjacent \htwo\
transitions, while the \sthree\ transition at 1190 \AA\ is adjacent to
and slightly contaminated by \ion{Si}{2} at 1190.416 \AA\ (a velocity
offset of $+52.4$ \kms\ with respect to \sthree).  We deal with these
contaminations in two separate ways.  We correct \sthree\
$\lambda$1190 for the presence of underlying absorption from
\ion{Si}{2} $\lambda$1190 using an apparent optical depth profile for
the \ion{Si}{2} derived from the \ion{Si}{2} transitions at 1193.290
and 1526.707 \AA.  In each case we scale the relative $\lambda
f$-values of the transitions involved in order to estimate the
contribution from the 1190.416 \AA\ transition of \ion{Si}{2} as a
function of velocity.  Using these two comparison lines, which differ
in strength by a factor of 3.4, leads to the same value for the
integrated \nav\ profile of \sthree\ $\lambda$1190: $\log
N_a(\mbox{\sthree}) = 14.66\pm0.04$.  

We assess the strength of the \htwo\ lines contaminating \sthree\
$\lambda$1012 following Howk et al. (2002), who discuss this issue for
contamination of \ovi\ absorption.  We assess the strength of the
\htwo\ lines contaminating \sthree\ $\lambda$1012 using lines of
similar $f$-values arising from the same rotational states.  In this
case, the derived column density is sensitive to the assumptions about
this correction and a somewhat uncertain continuum placement.  We
derive column densities from \sthree\ \wave{1012} that are
significantly lower than those derived from the STIS-observed \sthree\
\wave{1190}: $\log N_a(\mbox{\sthree}) \approx 14.53\pm0.05$ to
$14.59\pm0.04$ (statistical errors only), depending on the assumptions
about the continuum fit, details of the \htwo\ contamination
correction, and resolution of the \fuse\ channel from which the data
are adopted.  The smaller columns derived from this weaker line of
\sthree\ are smaller than those for the stronger line at 1190 \AA,
suggestive of some unresolved saturation in at least the \fuse\ data.
However, given the uncertainties in the measurements of \sthree\
\wave{1012}, it is difficult to directly compare the two values as we
did for \sii\ above.

On the other hand, the peak apparent optical depth of \sthree\ 1190 as
observed with the higher-resolution STIS instrument is only $\sim60\%$
that of \sii\ 1250 (including the underlying absorption from
\ion{Si}{2} 1190).  Even though they do not trace exactly the same
gas, it is unlikely the \sthree\ 1190 transition is saturated to the
extent of the \sii\ 1250 transition.  This, with the difference in
derived \nav\ values between the two transitions of \sthree, limits
the required saturation correction for the \sthree\ 1190 apparent
column density to be $<+0.07$ dex, the value adopted for \sii.  The
only other comparable probes we have of the low-velocity gas are the
\ion{Si}{4} 1393, 1402 \AA\ lines (Zech et al. 2008), which show
little to no apparent unresolved saturation for optical depths
$>2\times$ that of \sthree.  However, while their shape is similar to
that seen in \sthree, they are offset by $\sim -4$ \kms\ and trace a
somewhat different mixture of gas than \sthree.\footnote{This appears
  not to be due to a wavelength calibration uncertainty, as the
  \ion{Si}{4} high velocity cloud absorption seen in both transitions
  is consistent at velocities consistent with those seen in
  \ion{Si}{2} 1190.}  Profile fitting does not ultimately provide a
firm assessment of the possible saturation, as there is little
information on the intrinsic shape of the \sthree\ profile from other
transitions, and the \htwo\ contamination and unknown LSF that plague
the \fuse\ observations of the 1012 \AA\ transition make that approach
non-unique in our tests.  We proceed by adopting a saturation
correction of $+0.035$ dex and adding a systematic uncertainty of $\pm
0.035$ linearly to the statistical uncertainties.  We adopt this
uncertainty under the assumption that the correction should be $<0.07$
dex at the $2\sigma$ level.  Our adopted column density is thus $\log
N(\mbox{\sthree}) = 14.70\pm0.06$.


Several other metal species are listed in Table \ref{tab:columns},
which represent more straightforward measurements.
\fuse\ observations place stringent limits on the \ion{S}{4} column
densities through the 1062.664 \AA\ transition for both sight lines,
while we are also able to place limits on \ion{S}{6} toward \vz\ using
the doublet at 933.387 and 944.523 \AA.  For the \zng\ sight line, we
are not able to place meaningful limits on \ion{S}{6} given the strong
contamination from \htwo\ in the FUV.  Given the column of \ion{S}{4}
is well below that of \sthree\ toward \zng, we assume the \ion{S}{6}
column would be minimal toward this star as it is toward \vz.  Limits
to \ion{S}{1} $\lambda$1295.653 absorption from STIS show its column
density to be negligible along both sight lines.

\subsection{Neutral and Molecular Hydrogen Column Densities}

We derive the interstellar \HI\ column density of these two sight
lines by fitting the damping wings of the \lyalpha\ profile observed
by STIS.  For the sight line to \vz\ we adopt the \lyalpha\ derived
\HI\ column density from \HSS.  Here we determine the \HI\ column
density toward \zng\ in an identical manner.

The neutral hydrogen column in the direction of M~5 can be estimated
from \HI\ 21-cm emission observations.  As noted in Zech et
al. (2008), the column density derived from the publicly-available
Leiden-Argentine-Bonn (LAB) Survey (Kalberla et al. 2005) is $\log
N(\mbox{\HI}) = 20.56\pm0.17$.  This value is averaged over a
$36\arcmin$ beam approximately centered on M~5.  However, small-scale
structure within this large beam can cause this average value to be
different than that appropriate for comparison with the pencil-beam
measurements of metal ions toward \zng\ itself.  We adopt an error
following the recommendations of Wakker et al. (2001) for the use of
large beam \HI\ 21-cm observations for absorption line studies.

Figure 3 shows the STIS spectrum of the \lyalpha\ absorption line
toward \zng.  The broad \lyalpha\ absorption is a combination of
stellar atmospheric and foreground ISM absorption.  We remove the
contaminating stellar absorption by using the stellar atmosphere model
described in Zech et al. (2008) to normalize the data during our
fitting procedure.  The model atmosphere was calculated with TLUSTY
(Hubeny \& Lanz 1995) assuming an effective temperature $T_{eff} =
45,000$ K, with $\log g = 4.48$, and an atmosphere consisting of 99\%
He by number (W.V. Dixon, 2007, private communication) with a mix of
abundances specified in Zech et al. (2008).  The adopted stellar
continuum is shown as the blue line in Figure 3.  We use a
second-order Legendre polynomial to match the model atmosphere to the
STIS spectrum.  Ideally the model atmosphere should provide a very
good estimate of the distribution of flux with wavelength.  In
practice the models deviate slightly from the spectral energy
distribution in the data.  Small discrepancies between the model and
observations are due to large-scale calibration and small-scale order
combination uncertainties in the data and uncertainties in the
large-scale flux distribution of the model itself.  The polynomial
parameters are treated as free parameters during the fitting process
and contribute appropriately to the error budget.

The best-fit interstellar \HI\ column density from our modeling of
\lyalpha\ is $\log N(\mbox{\HI}) = 20.47\pm0.02$.  The best-fit
profile is shown as the red line in Figure 3.  This error estimate is
dominated by uncertainties the adopted stellar properties and their
affect on the stellar models, which we explore in the fitting process
(c.f., Sonneborn et al.  2002).

The molecular hydrogen content for both sight lines is completely
negligible compared with the \HI\ and \HII\ columns.  The sight line
to \vz\ shows no detectable \htwo\ absorption (Howk et al. 2003), with
$\log N(\mbox{\htwo}) < 14.35\ (3\sigma)$ summed over $J\le 3$.  This
implies a molecular hydrogen fraction $\log f(\mbox{\htwo}) \equiv
\log 2N(\mbox{\htwo}) /[N(\mbox{\HI}) + 2N(\mbox{\htwo})] < -5.3$.
Toward \zng, our \fuse\ data show a forest of moderate strength \htwo\
transitions.  However, none show damping wings, which severely limits
the column densities.  Deriving the precise column density is not
crucial for this sight line, and we use the lack of damping wings and
other considerations to limit the column: $\log N(\mbox{\htwo}) <
18.0\ (3\sigma)$ integrated over $J\le5$ states.  This yields $\log
f(\mbox{\htwo}) < -2.4$.  The \htwo\ contribution to the total
hydrogen column along both sight lines is negligible.

\subsection{Radio Pulsar Dispersion Measures and the Ionized Hydrogen
  Column Densities}
\label{sec:pulsars}

The electron column densities toward M~3 and M~5 are derived from the
dispersion measures toward millisecond pulsars in each cluster.
Following \HSS\ we average the dispersion measure for three M~3
pulsars (M~3A, M~3B, and M~3D), giving $\langle \dm \rangle _{{\rm
    M~3}} = 26.33\pm 0.15$ pc \percc\ (standard deviation).  We do not
include the unconfirmed pulsar M~3C in this average, although it gives
a consistent DM (Hessels et al.  2007).  The uncertainties in these
\dm\ measurements are very small, typically $\sim0.1$ pc \percc\ or
better.  This average dispersion measure is equivalent to an electron
column density of $\log \Nelectron = 19.91\pm0.01$ in typical units
(cm$^{-2}$).  For M~5 we average the $DM$ for the five known
millisecond pulsars (Hessels et al. 2007, Freire et al. 2008) giving
$\langle \dm \rangle _{{\rm M~5}} = 29.5\pm 0.3$ pc \percc\ (standard
deviation) or $\log \Nelectron = 19.96\pm0.01$.  These are the total
electron columns, including contributions from the WIM and the HIM
along these sight lines.

The application of Equation \ref{eqn:himelectrons1} to the sight lines
in this work yields $\log \Nelectron\subhim\sim 19.08$ and $19.00$
\column\ for the \vz\ and \zng\ sight lines, respectively (see Table
\ref{tab:results}).  We assume \ovi\ column densities from Table
\ref{tab:columns}.  The result for \vz\ is slightly different than
that reported in Howk et al. (2006) given the different approach to
assessing the contribution from the hottest HIM component,
$N(\mbox{\HII})_6$, but the difference is small enough that it
produces a negligible change in the final results.  For example, the
$A({\rm S})$ reported here is different by only $\sim0.01$ dex
compared with the earlier value.  Adopting the Howk et al. methodology
for the \zng\ sight line, however, would produce results that differ
by $\sim0.1$ dex given the greater importance of lower $z$-height gas
along that sight line.

\section{Results}
\label{sec:analysis}


The final results of our WIM analysis are presented in Tables
\ref{tab:results} and \ref{tab:fractions}.  Table \ref{tab:results}
gives the physical parameters derived for each of the sight lines,
including the hot gas columns, WIM fractions, and $A({\rm S})$.  Table
\ref{tab:fractions} presents the final ionization fractions for S$^0$,
S$^{+}$ (see below), S$^{+2}$, S$^{+3}$, and S$^{+5}$ in the WIM for
the two sight lines.  To derive the ionization fractions, we adopt the
S abundance toward M~3, which is much better determined than that
toward M~5 and equivalent to the solar system meteoritic
abundance. Sulfur does not seem to be depleted by large amounts into
dust grains (again evidenced by the solar-like abundance toward M~3),
so it is reasonable to assume $A({\rm S})$ is roughly constant in the
solar neighborhood. While the M~5 sight line is toward the inner
Galaxy, the majority of the absorption likely occurs within the first
$z<1$ kpc given the scale height of the WIM, and thus relatively close
to the Sun.

The two sight lines in this study may trace different conditions in
the WIM.  The high-latitude cluster M~3 is close to the Galactic north
pole and probes the WIM in a column above the sun.  On the other hand,
M~5 lies toward the inner Galaxy at $b \sim +45^\circ$, probing WIM
gas somewhat interior to the solar circle.  As shown in Table
\ref{tab:results}, the mean fractions of ionized gas along the sight
lines differ by a factor of $\sim2$.  The sight line to M~3 has a
total ionized gas fraction $N(\mbox{\HII})/N({\rm H}) =
0.45\pm0.09$,\footnote{The values quoted here for the M~3 sight line
  differ slightly from those of Howk et al. (2006) due to the
  difference in the treatment of the HIM contribution.} whereas that
toward M~5 is only $0.19\pm0.03$.  This is likely due in large part to
the differing paths through low-\z\ gas associated with the denser,
more neutral thin disk of the Galaxy.  The M~5 sight line probes a
proportionally-larger contribution from the low-\z\ gas.

Within the warm ionized gas, we find ionization fractions
$x(\mbox{\spp}) = 0.33\pm0.07$ for M~3 versus $0.47\pm0.09$ for M~5
(Table \ref{tab:fractions}).\footnote{Utilizing the directly measured
  value for $A({\rm S})$ toward M~5 yields $x(\mbox{\spp}) =
  0.67\pm0.19$.}  Here we have transformed from $x(\mbox{\spp})/x({\rm
  H}^+)$ assuming a hydrogen ionization fraction $x({\rm H}^+) =
0.95\pm0.05$ to represent the WIM, which generally has $x({\rm H}^+) >
0.9$ (Reynolds 1989, Reynolds et al. 1998a, Hausen et al. 2002,
Haffner et al. 2009).  These ionization fractions are column
density-weighted averages over the entire sight lines.  We do not
detect \sfour\ absorption along either sight line, and this absence
limits the ionization fraction $x({\rm S}^{+3}) < 0.04$ ($3\sigma$)
for both. The ionization fractions of S$^{0}$ in the WIM are very low,
as expected: $x({\rm S}^0) < 0.004$ ($3\sigma$) along both sight
lines.

The ionization fraction of S$^+$ in the WIM cannot be directly
measured through the column density of \sii, since that ion contains
significant contributions both from the WNM and WIM (with the WNM
dominating the column).  However, none of the ionization states higher
than \spp\ contribute significantly to the total, with very stringent
$3\sigma$ limits to the ionization fractions of ${\rm S}^{+3}$ and
${\rm S}^{+5}$.\footnote{While we do not have a probe of ${\rm
    S}^{+4}$, the lack of significant quantities of any other species
  with ionization energies $\ga35$ eV (in the WIM or the hot gas up to
  $\ga150$ eV; see below), we do not expect S$^{+4}$ to be present in
  significant amounts.}  Given no higher ionization states are
present, the ionization fractions of S$^+$ and \spp must sum to nearly
unity, i.e., $x({\rm S}^+)+x({\rm S}^{+2}) \approx 1$.  We use this to
estimate the ionization fractions of S$^+$ within the WIM along these
sight lines: $x({\rm S}^+) = 0.67\pm0.07$ and $0.53\pm0.09$ for M~3
and M~5, respectively (allowing for a $1\sigma$ contribution from
S$^{+3}$ in the error budget).

The distribution of S ion fractions in the WIM as a function of
ionization energy is shown in the top panel of Figure
\ref{fig:ionfracs1} for S$^0$ through S$^{+5}$ (excluding S$^{+4}$),
with values for the M~3 and M~5 sight lines shown in black and red,
respectively.  The horizontal bars for each ion span the creation to
destruction energies for each ionization state.  On the whole the
results from the two sight lines are in good agreement.  There is a
strong peak in the ion fractions for energies between $\sim10$ and 35
eV, with very little at higher energies.  On the whole the two sight
lines give a consistent picture of the ionization distribution.  The
M~5 sight line may slightly favor \spp\ compared with that to M~3, but
any variations are at less than $2\sigma$.


Independent information on the metal ion fractions in the WIM can be
derived from emission line observations.  Results derived from such
studies are compared with ours in Figure~\ref{fig:ionfracs2}.  Haffner
et al. (2009) give a recent review of WIM emission line observations
and physics (both for the Galactic WIM and the diffuse ionized gas in
external galaxies).  They combine the observations of Haffner et
al. (1999) and Madsen et al. (2006), who used WHAM to observe
forbidden metal line emission from the diffuse Galactic WIM.  Their
observations of the ratio [\sii]/\halpha\ provide a measure of $x({\rm
  S}^+)$ when the temperature can be estimated with observations of
[\ion{N}{2}]/\halpha\ (Haffner et al. 1999).  As summarized in Haffner
et al. (2009), the WHAM results give $x({\rm S}^+) \sim 0.3$ to 0.7,
with the majority of the sight lines seemingly above 0.5.  The results
for the two sight lines studied here, giving $x({\rm S}^+) \sim 0.5$
and 0.7, are completely consistent with the WHAM estimates (Figure
\ref{fig:ionfracs2}).

The lack of \sfour\ and higher ion absorption is also consistent with
WHAM observations of forbidden [\othree] emission from the WIM, which
show $x({\rm O}^{+2}) \la 0.1$ in all cases and $<0.05$ for a majority
of the diffuse gas (Madsen et al. 2006).  \othree\ probes a range of
ionization energies 35.12 to 54.93 eV, similar to those probed by
S$^{+3}$ (see Figure \ref{fig:ionfracs2}).  Similarly, observations of
\ion{He}{1} recombination radiation limit the ionization fraction of
singly-ionized helium.  Observations find ionization fractions of
singly-ionized helium as high as $x({\rm He}^+) \la 0.6$, although
many sight lines have $x({\rm He}^+)\la 0.3$ or so (Madsen et
al. 2006; Reynolds \& Tufte 1995).  Given the lack of higher
ionization states of S from our results, the remainder of the He is
likely to be neutral (see \S \ref{sec:discussion}).

There seems to be a significant amount of variation in the O$^{+2}$
and He$^+$ ionization fractions with location in the Galaxy.  The
highest ionization fractions from Madsen et al. (2006) in particular
seem to be found associated with energetic regions very near the disk
or in unusual filaments.  The diffuse WIM and even supershells tend to
trace the lower end of these ionization fractions.  Figure
\ref{fig:ionfracs2} shows the full range of observed values, from the
lowest upper limits (shown with downward pointing arrows) to the more
extreme, higher ionization sight lines (Madsen et al. 2006).  However,
on the whole the emission line observations paint a similar picture to
our absorption line study: ions requiring $>35$ eV for production make
a relatively small contribution to the total.


While we have concentrated to this point on the ionization fractions
within the WIM, we also have an opportunity along these sight lines to
consider the ionization fractions of a broad range of ions in the
total ionized gas of the Milky Way (i.e., WIM+HIM).  Here we write the
total column density of \HII\ as
\begin{equation}
\label{eqn:totalhii}
N(\mbox{\HII}) = [(1-f\subhim) \eta\subwim + f\subhim \eta\subhim] \Nelectron,
\end{equation}
where $f\subhim = N(\mbox{\HII})\subhim / N(\mbox{\HII})$ is the
fraction of the \HII\ column associated with the HIM.  This value is
derived from our estimates of the \HII\ column associated with the hot
ISM as discussed in \S \ref{sec:HIM}.  While the HIM fraction is
model-dependent and poorly constrained observationally, it is only
important in this case so far as it affects the mean value of $\eta$,
which is given in the term in square brackets in Equation
\ref{eqn:totalhii}.  We assume 50\% uncertainties in
$N(\mbox{\HII})\subhim$, although its precise value does not affect
the total \HII\ column substantially.  With this total \HII\ column in
hand, we write the total ionization fraction of an ion $X^j$ as
\begin{equation}
x(X^j)_{total} = \frac{N(X^j)}{N(\mbox{\HII})} A(X)^{-1},
\end{equation}
equivalent to Equation 2.  Figure \ref{fig:totalionfracs1} shows the
results of this approach as applied to the S ions as well as the high
ions Si$^{+3}$, C$^{+3}$, N$^{+4}$, O$^{+5}$.  The values are
summarized in Table \ref{tab:totalionfracs}.  We adopt solar system
abundances for these ions.  For CNO, these abundances are probably
reasonable, since those elements are not depleted by more than a
factor of 2 (perhaps less in the hot ISM).  Incorporation of Si into
the dust phase could lower the gas-phase abundance of Si quite a bit,
probably even in the WIM (Howk \& Savage 1999).  Thus, the ion
fraction of Si$^{+3}$ could be somewhat higher than shown, perhaps by
up to a factor of a few.

\section{Discussion}
\label{sec:discussion}

We have presented direct measures of the ionization fractions of
several ions of S in the WIM of the Milky Way (Figure
\ref{fig:ionfracs1}), as well as of the ionization fractions of several
metal ions in the integrated (warm and hot) ionized gas of the Milky
Way (Figure \ref{fig:totalionfracs1}).  These figures represent
``maps'' of the preferred ionization energies in the ionized gas of
the Milky Way, showing at what energies we expect to find significant
ionization fractions, albeit maps that are slightly skewed by atomic
physics (through varying ionization cross sections and recombination
coefficients).  In both cases, we find only those ions requiring $<35$
eV for their production are present at more than the few percent level
up to energies $\sim150$ eV.  This constrains the sources of
ionization for the WIM and the relative mass contributions as a
function of temperature for the hotter, collisionally ionized gas (see
below).

The WIM shows no absorption from S ions above \spp.  The limits on
S$^{+3}$ and higher absorption are meaningful for several reasons.
First, the lack of high ionization S absorption suggests that the only
ionization states of S with significant ionization fractions in the
WIM are S$^+$ and \spp, allowing us to estimate the S$^+$ ionization
fraction (see \S \ref{sec:analysis} and Table \ref{tab:fractions}).
The S$^+$ ionization fractions in the WIM along these two sight lines
is quite a bit higher than found in classical O-star \HII\ regions,
which typically have $x({\rm S}^+) \approx 0.25$ (Haffner et
al. 2009), with \spp\ being a dominant ion state in those regions.
Thus, the WIM has a generally lower ionization state than these
\HII\ regions, as noted previously (Haffner et al. 1999, Madsen et
al. 2006).  

Second, the lack of S$^{+3}$ (and S$^{+5}$ toward \vz) severely limits
the amount of high ionization gas produced by the ionization sources
for the WIM.  The energies required to produce and destroy S$^+$ are
10.36 eV and 23.33 eV; for \spp\ these are 23.33 eV and 34.83 eV.
Thus, there is not a significant contribution to the S budget from any
ions requiring $\ga35$ eV for their production.  Essentially all of
the S in the WIM is in either S$^+$ or \spp.  While the ultimate
sources of WIM ionization are not precisely known, our measurements
require they not produce significant amounts of S ions higher than
S$^{+2}$.  This is of particular importance for understanding the
ionization state of He.  The energy required to ionize S$^+$ is close
to that required to ionize He$^0$ (23.33 eV versus 24.59 eV), and the
ionization potential of S$^{+2}$ is well below that required to ionize
He$^{+}$ (47.22 eV versus 54.42 eV).  Thus, the lack of S$^{+3}$
(probing ionization energies 34.79 to 47.22 eV) thus argues that
virutally no He$^{+2}$ (probing energies 54.42 eV and above) can be
present in the WIM.

This conclusion about the lack of He$^{+2}$ is at odds with the work
of Arabadjis \& Bregman (1999), who claim that ``there is little room
for warm ionized gas of moderate ionization state.''  Based on an
analysis of the ISM opacity to X-ray emission from distant galaxy
clusters, these authors conclude that the fraction of neutral He in
the WIM must be very small, with all of the He in either He$^+$ or
He$^{+2}$.  Since the emission line observations of \ion{He}{1}
recombination emission only provided a measure of the He$^+$ ion
fraction, this conclusion could not immediately be ruled out by the
existing observations.  However, our results for the ionization
fractions of S$^{+3}$ and S$^{+5}$ coupled with measurements of
[\othree] emission (Madsen et al. 2006, Reynolds 1985b) show that ions
requiring $\ga 35$ eV for creation are trace ions in the WIM.  Our
limits on \sfour\ give $x({\rm S}^{+3})<0.04$ ($3\sigma$) for both
sight lines.  This ion is destroyed at an energy of $\sim47$ eV, well
below the energy required to produce He$^+$, $\sim55$ eV.  The
ionization cross sections and recombination coefficients of S and He
are not so different that a very high ionization fraction of He$^{+2}$
could coexist with such low ionization states of S$^{+3}$ (e.g.,
Sembach et al. 2000).


Several models have been developed in an attempt to match the metal
ion emission from the WIM.  Most of these are 1D models relying on
photons from OB stars to provide the ionization (e.g., Mathis 1986,
2000; Domgorgen \& Mathis 1994, Sembach et al. 2000, Elwert \& Dettmar
2005).  These are necessarily simplified models that often assume a
single temperature stellar atmosphere and very basic geometry, but
include a great deal of atomic physics.  Because many of these are
largely concerned with matching the emission lines, which are
temperature dependent, they do not all discuss directly the metal ion
fractions in a way that is appropriate for comparison with the
absorption line-derived results presented in this work.  Sembach et
al. (2000) attempted to model the WIM ionization using a series of 1D
models within Cloudy.  They specifically tailored their models to
assess the impact of the WIM on absorption line measurements.  Their
recommended composite model gives $x({\rm S}^{+})=0.81$ and $x({\rm
  S}^{+2})=0.18$.  The \spp\ fraction, in particular, is lower by
factors of 2 to 3 than our observations imply.  Indeed, Howk et
al. (2006) noted that applying the Sembach et al. (2000) models to the
\vz\ sight line did not produce results consistent with the full
abundance determination for sulfur (underestimating the total S
abundance by -0.4 dex).  The sign of this result implies the ratio
$x({\rm S}^+)/x({\rm S}^{+2})$ in the Sembach et al. models is too
large, consistent with our determinations.


More recent models have considered the modification of the radiation
field as it propagates through the ISM (e.g., Wood \& Mathis 2004,
Giammanco et al. 2004).  The absorption of some radiation as it passes
through the ISM will remove photons just above 1 Rydberg,
``hardening'' the H-ionizing photons as $E\approx1$ Ryd photons are
removed while higher energy photons see less opacity.  However, this
will tend to increase the S$^+$ ion fractions compared with \spp,
since the former is created by unabsorbed $E<1$ Ryd photons while the
radiation field is diminished at the energy required to create
\spp\ (Hoopes \& Walterbos 2003).  However, this is true only when
keeping all else constant.  Elwert \& Dettmar (2005) attempted to
model the run of [\sii] emission with height above the Perseus Arm
observed by Haffner et al. (1999), following the propagation of
radiation through the ISM with height.  However, while their models do
produce significantly higher S$^+$ ion fractions well above the plane
(and hence are able to match the run of [\ion{S}{2}] emission at
heights $z>1$ kpc above the Perseus Arm), they produce much lower
S$^+$ ion fractions at low heights.  Their models have $x({\rm S}^+)=
0.2$ to $0.5$, with the ion fraction increasing with height up to
$z\sim2$ kpc (see their Figure 5). Integrating their S$^+$ ionization
fractions vertically through an assumed exponential WIM distribution
with a 1 kpc scale height (e.g., Haffner et al. 1999) gives a mean
$\langle x({\rm S}^+)\rangle = 0.28$, lower by a factor of 2 than the
value implied by our observations.  A lower latitude sight line
through such a model (approximating that to \zng) would yield an even
lower mean ionization fraction.  The majority of the S in these models
is likely in S$^{+2}$ at $z\la1$ kpc, which dominates the column
density of the distribution.  The \spp\ ion fractions in these models
are significantly higher than our measurements imply.

We are not aware of significant constraints on the \spp\ ionization
fraction from emission line observations that could have been used to
guide these earlier models.  The emission line diagnostics also have
an additional dependence on temperature that can mask mismatches
between the model and WIM ionization fractions.  Our measurements
provide complementary constraints for models of the WIM. The causes of
the discrepancies between the two models discussed above and the
ionization fractions measured here are unknown at this point, but may
include inappropriate choice of ionizing spectra, effects associated
with the 3D structure of the gas and radiative transfer, and the lack
of appropriate heating sources (which affects the match with emission
line diagnostics, but not the absorption line diagnostics here).  The
ionizing spectrum is likely to be quite complex with multiple types of
sources contributing.  While it is almost certainly dominated by O
star radiation (Haffner et al. 2009), that radiation is processed by
intervening absorption and there may be contributions from other
sources such as stellar remnants (e.g., Bregman \& Harrington 1986,
Rand et al. 2011) and cooling radiation from hot gas and
transition-temperature interfaces (Slavin et al. 2000).

We note that the conditions in the WIM near the sun may be somewhat
different than those seen in other galaxies, where obserations of the
diffuse ionized gas (DIG) often trace layers brighter than the WIM of
the Milky Way (Haffner et al. 2009).  In particular, Rand et
al. (2011) have recently used {\em Spitzer} observations of the
[\ion{Ne}{3}]/[\ion{Ne}{2}] ratio to show the ratio of
Ne$^{+2}$/Ne$^+$ is increasing with height above the plane.  Because
these emission lines are not very sensitive to excitation in the way
the optical forbidden lines can be, this is a robust measure of the
relative importance of these Ne ions.  Observations of [\ion{O}{3}]
emission also find some galaxies show higher O$^{+2}$ ionization
fractions than implied by the WHAM measurements of the Milky Way.
Collins \& Rand (2001) study, for example, showed that O$^{+2}$ could
be the dominant ionization state of O in the extraplanar DIG of NGC
5775 and UGC 10288.  One concern with comparing the Milky Way and
these other galaxies is the difference in the properties of the
observed DIG layers compared with the Milky Way's WIM.  The former
tend to be quite bright compared with the WIM and are observed at
projected radii well within the solar circle in most cases.  However,
this is suggestive that the results found here may not apply across
all galaxies and even over all positions within the Milky Way.

The ionization fractions shown in Figure \ref{fig:totalionfracs1}
demonstrate the importance of the WIM in the strong peak between
$\sim10$ and 35 eV, but they also shed light on the conditions in the
hotter ionized gas of the Milky Way.  With the exception of S$^+$ and
\spp, none of the ions in Figure \ref{fig:totalionfracs1} exceeds an
ionization fraction of 0.05.  The very low ionization fractions of the
high ions Si$^{+3}$, C$^{+3}$, N$^{+4}$, O$^{+5}$ have two causes.
First, none of these Li-like ions (or Na-like in the case of
\ion{Si}{4}) is expected to be the dominant ionization state at any
temperature, having peak ionization fractions of $\approx0.25$ (Gnat
\& Sternberg 2007).  However, the values seen in Figure
\ref{fig:totalionfracs1} are an order of magnitude or more lower than
this.  The low ionization fractions for these high ions reflect the
relatively small contribution of gas in the transition temperature
regime with $T\sim 7 \times 10^4 - 4\times10^5$ K to the total ionized
gas content of the Milky Way.  This is understandable: gas at these
temperatures is at the peak of the interstellar cooling curve and
radiates its energy very quickly, thereby cooling very quickly through
this temperature regime.  Thus, while the total mass budget of such
transition temperature gas is not large compared with the WIM or the
hotter HIM (see below), it is very important energetically.  The mass
flux or cooling flux through this regime is likely to be quite
important.  In fact, the ion fractions for these high ions include
contributions from such transition-temperature gas as well as the
$\sim40\%$ of such gas that is at $T<7\times10^4$ K, representing
matter that has already cooled and is out of ionization equilibrium or
gas that is photoionized by the radiatively cooling HIM (Lehner et
al. 2011).

Figure \ref{fig:totalionfracs1} hints that there are likely to be two
peaks in the total ionization fractions in the ionized gas of the
Milky Way.  One is associated with the WIM and traced by ions that
probe the energy range $E\sim10$ to $35$ eV, as demonstrated by the
large ionization fractions of S$^+$ and \spp.  This gas is
predominantly photoionized and warm.  The second peak is unseen in
this plot, since we have no ions probing energies $E\ga 140$ eV.  The
gas associated with the hot ISM at $T\ga 4\times10^5$ K (above that
typically probed by \ovi) will be traced in O by O$^{+6}$ and
O$^{+7}$; such gas will be traced in S by S$^{+6}$ and higher (e.g.,
Sutherland \& Dopita 1993, Gnat \& Sternberg 2007).  While we cannot
directly probe these ionization states along these sight lines, we can
strongly limit their combined contribution.  Summing the ionization
fractions of S$^+$ and \spp, we have $x({\rm S}^+)+x(\mbox{\spp})
\approx 0.80\pm0.10$ along both the M~3 and M~5 sight lines.  Thus,
the combined fractions of all other ionization states represent only
$\approx(20\pm10)\%$ of the total S in these directions.  The second
peak in the ionization fractions of S ions should trace the high
temperature HIM at a few hundred eV and have a magnitude of $\la0.2$.  
Figure \ref{fig:totalionfracs2} shows a plot of our derived ionization
fractions supplemented by a calculation of sulfur ionization fractions
for a hot, collisionally-ionized phase with $T=2\times10^6$ K (e.g.,
Yao \& Wang 2006).  The hot phase ion fractions are from a Cloudy
(Ferland et al. 1998) calculation assuming collisional ionization
equilibrium (CIE) and have been normalized to sum to 20\% of the total
(although the total for these high energies may be somewhat less).
Thus, this figure shows a map of the expected energies at which one
expects to find ions within the ionized medium of the Milky Way.  The
distribution in the highest ions depends on the assumed temperature of
the HIM, but the other ions now have firmly measured ionization
fractions along these two sight lines.

The very low ion fractions of the high ions observable in the UV
(esp. O$^{+5}$) have implications beyond the Galaxy.  The strong
\ovi\ doublet is among the best tracers of highly-ionized gas in
galaxy halos over all redshifts (Prochaska et al. 2011, Tumlinson et
al. 2011).  Studies of this sort that attempt to make estimates of the
mass of gaseous galaxy halos seen in absorption, correcting the
\ovi\ columns based upon an assumed ionization fraction from
collisional ionization models (e.g., Gnat \& Sternberg 2007).  In
these cases, the typical ionization fraction adopted is $x({\rm
  O}^{+5}) \la 0.2$ (Gnat \& Sternberg 2007, Sutherland \& Dopita
1993).  This may be the ionization fraction in the \ovi -bearing gas
at the temperatures where \ovi\ peaks in abundance, and this gas
represents a small fraction of the total, as we have discussed above.
However, if one wishes to use \ovi\ to calculate the total mass of
ionized gas or the mass of hot ionized gas (e.g., $T\ga10^5$ K in this
case), our results suggest some caution may need to be used, asa much
lower mean ionization fraction is suggested for our two sight lines.
In the two sight lines studied here, we find $x({\rm O}^{+5}) \la
0.01$ when considering the warm and hot ionized gas together.  The
\ovi\ along these sight lines represents $\la5\%$ of the oxygen in the
hot gas.  The applicability of our results, which probe the thick disk
of the Milky Way within $z\la10$ kpc of the midplane, to much
different environments is unclear.  Many studies in which such
corrections are important probe galaxy halos at several tens of kpc
impact parameter from galaxies (e.g., Prochaska et al. 2011, Tumlinson
et al. 2011) or probe the outskirts of HVCs (e.g., Fox et al. 2006,
Sembach et al. 2003), which may have different ionization conditions
than the gas in the Milky Way thick disk.

\section{Summary}
\label{sec:summary}

We have presented a method for studying the ion fractions of metal
ions in the ionized gas of the Galaxy, expanding upon the method
presented by Howk et al. (2006) for measuring elemental abundances.
We make use of unique sight lines toward globular clusters
containing both UV bright stars (for measuring the metal ions) and
pulsars (for measuring the electron column density).  We apply this
method to study the ionization of the ionized gas toward \vz\ and
\zng, both of which probe extended paths through the Galactic thick
disk or halo, providing estimates for the ionization fractions of
S$^0$, S$^{+}$, S$^{+2}$, S$^{+3}$, and S$^{+5}$ in the WIM.  We also
assess the ionization fractions integrated through the warm and hot
ISM of the Milky Way.

Our principal conclusions based on this analysis are as follows.

\begin{enumerate}

\item The only ions with significant ionization fractions in the WIM
  are those with ionization potentials $\la35$ eV.  We find \spp\
  makes up a substantial portion of the WIM sulfur, with ionization
  fractions $x({\rm S}^{+2})=0.33\pm0.07$ and $0.47\pm0.09$ toward M~3
  and M~5, respectively.

\item We limit the contribution of S$^{+3}$ and higher ionization
  states of sulfur, with $3\sigma$ upper limits $x({\rm S}^{+3})<0.06$
  for both of our sight lines.  Given the lack of higher ionization
  states in the WIM, the ionization fractions of S$^+$ and \spp\
  represent the vast majority of the WIM sulfur.  Assuming these add
  to unity, we derive $x({\rm S}^{+})=0.67\pm0.07$ and $0.53\pm0.09$
  toward M~3 and M~5, respectively.  These numbers are in good
  agreement with emission-line derived S$^+$ ion fractions (Haffner et
  al. 1999, Madsen et al. 2006).

\item Existing, simplified photoionization models for the WIM produce
  S ion fractions that can differ by factors of 2 from our estimates,
  with various models under- or over-producing \spp.  These models are
  largely constructed to match emission line measurements, which
  provide access to a narrower range of ions than our measurements.
  The lack of higher ionization states in the WIM rules out the
  conclusion of Arabadjis \& Bregman (1999) that much of the He in the
  WIM is in the form of He$^{+2}$ and limits the hardness of the
  radiation field responsible for ionizing the WIM.

\item We show that the ionized gas column density (and hence mass)
  along these two sight lines is dominated by warm photoionized gas
  ($T\sim10^4$ K) favoring ionization states with ionization energies
  $E\sim10$ to 35 eV and a hot ionized phase ($T\ga 4\times10^5$ K)
  favoring energies of a few hundred eV.  The warm phase contains
  $\sim80$\% of the total ionized gas column, and the hot phase
  $<20\%$.  The important tracer \ovi\ probes only a small fraction of
  the ionized gas along the sight line with $x({\rm O}^{+5}) < 0.01$.

\end{enumerate}

\acknowledgements

Financial support was provided through NASA grants HST-GO-9150.03-A
and HST-GO-9410.03-A through Space Telescope Science Institute, which
is operated by the Association of Universities for Research in
Astronomy, Inc., under NASA contract NAS5-26555.  We thank N. Lehner,
J.X. Prochaska, R. Rand, and J. Tumlinson for useful comments and
discussions.



\begin{figure}
\epsscale{0.9}
\plotone{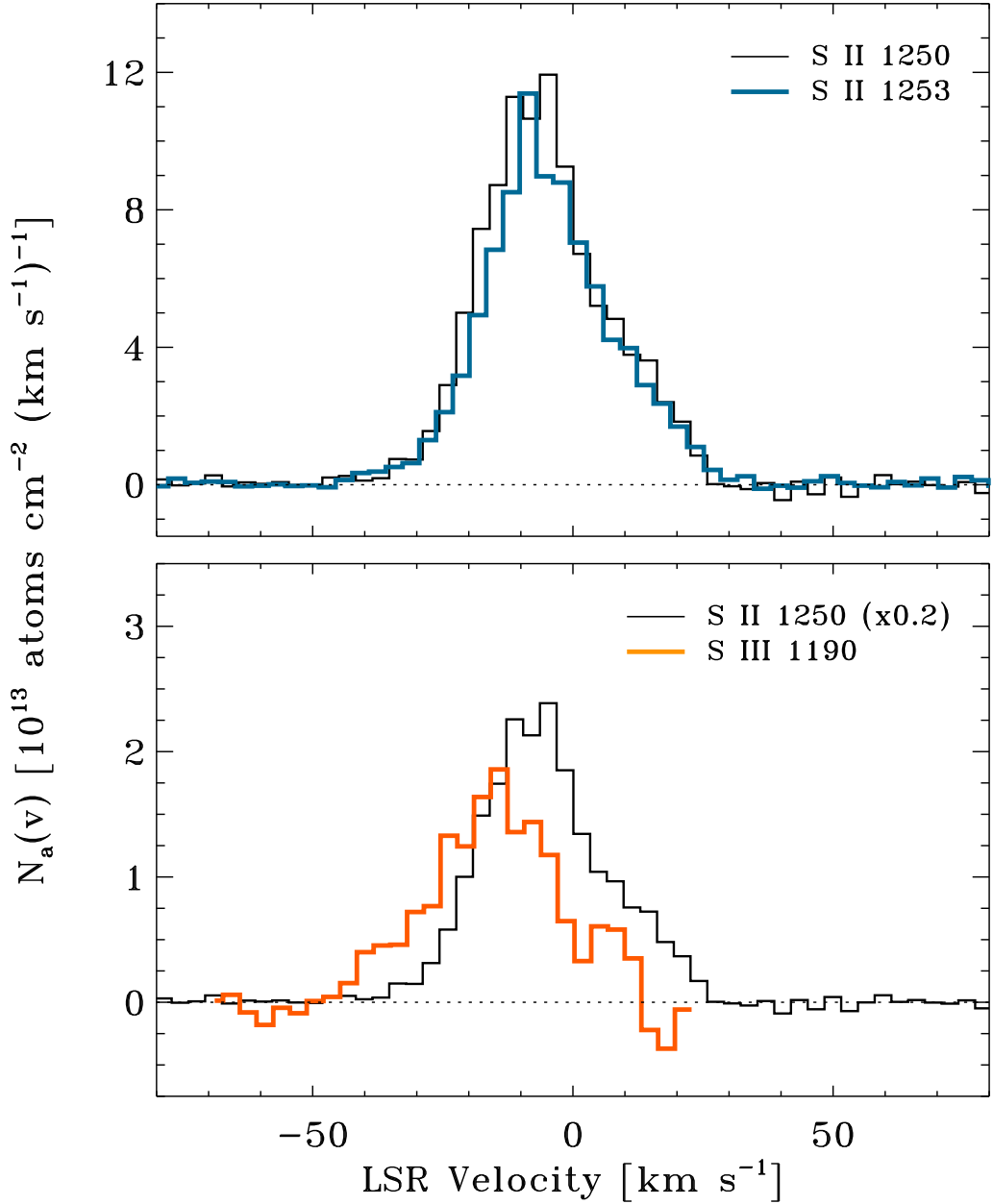}
\caption{Comparison of the apparent column density profiles for
  \sii\ \wave $1250.584$, $1253.811$, and \sthree\ \wave $1190.208$
  from STIS E140M observations.  The top panel is a comparison of the
  two \sii\ profiles.  The discrepancy between the two profiles is due
  to the presence of unresolved saturation. The lower panel compares
  the \sii\ line at $1250.54$ \AA\ with the \sthree\ line at
  $1190.208$ \AA. The \sthree\ profile has been ``cleaned'' of
  \ion{Si}{2} contamination. \label{fig:comparison}}
\end{figure}


\begin{figure}
\epsscale{0.9}
\plotone{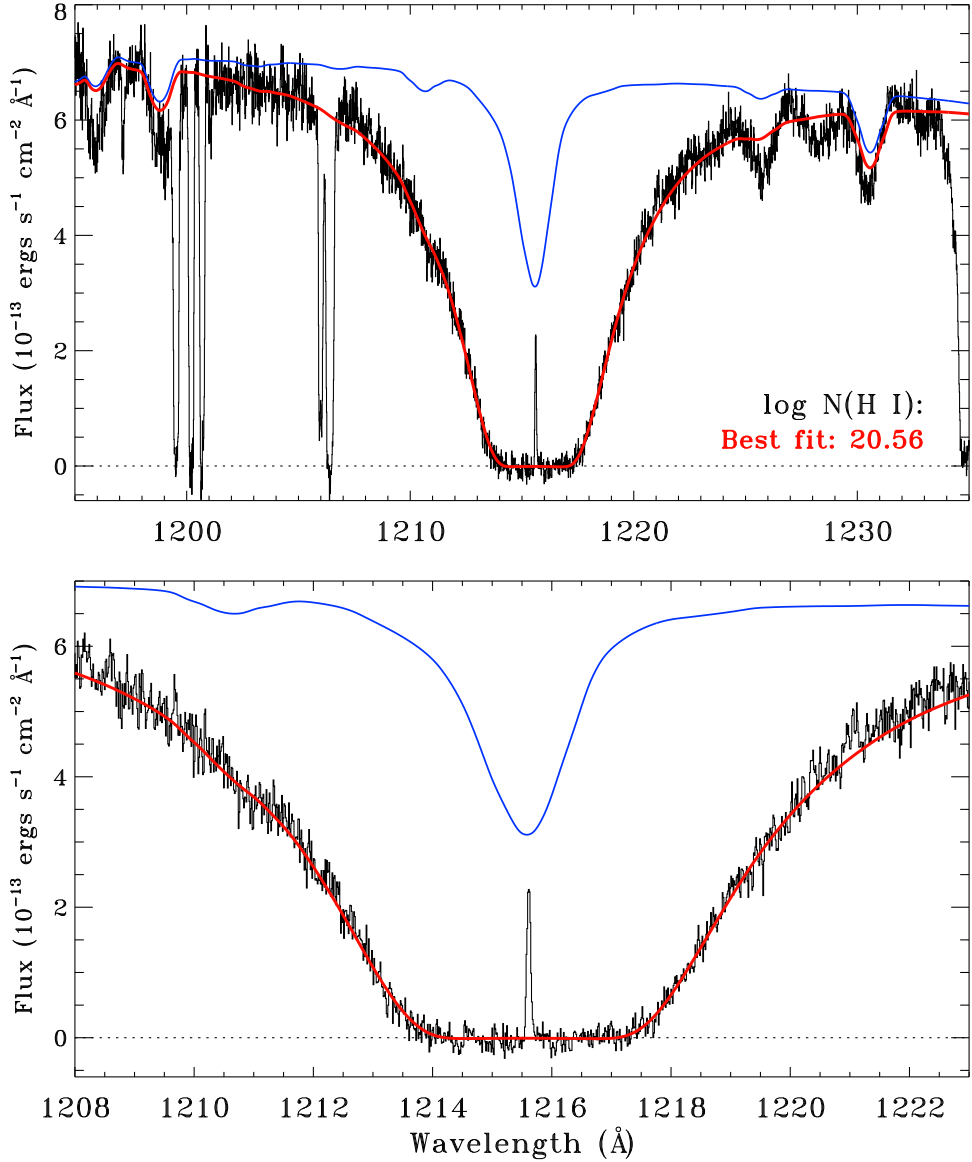}
\caption{Two views of the STIS E140M spectrum of the \lya\ absorption
  profile toward \zng. Both stellar and interstellar absorption
  contribute to this profile. The estimated stellar profile is shown
  as the thin blue line and has been shifted to \vlsr\ = +65.7
  \kms\ to match the observed positions of stellar absorption lines in
  the STIS spectrum. The best fit to the interstellar and stellar
  absorption profile is shown in red, corresponding to an interstellar
  column density $\log N(\mbox{\HI}) = 20.47\pm0.02$.  The sharp line
  in the center of the \lya\ absorption trough is geocoronal
  emission. \label{fig:lyalpha}}
\end{figure}


\begin{figure}
\epsscale{0.9}
\plotone{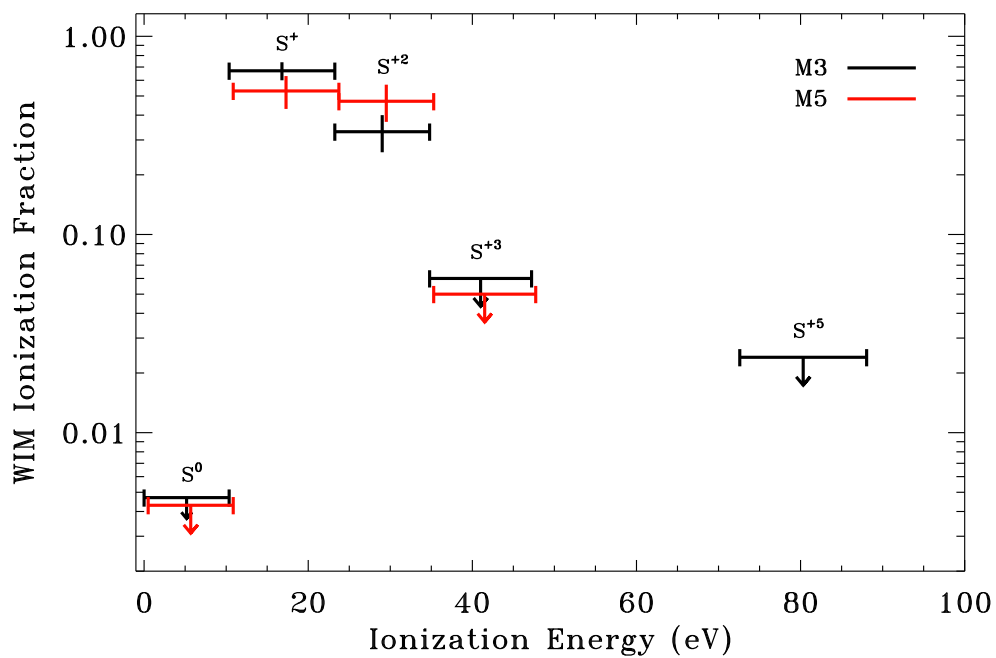}
\caption{The ionization fractions, $x(X^j)$, of S ions in the WIM as a
  function of ionization energies for the M~3 (black) and M~5 (red)
  sight lines.  The horizontal extent of the bars for each ion span
  the range of energies required to create it, from the next lower
  ionization state, and destroy it, creating the next higher
  ionization state (i.e., the ionization energies).  The M~5
  ionization energies are offset by 0.5 eV for clarity.  \label{fig:ionfracs1}}
\end{figure}

\begin{figure}
\epsscale{0.9}
\plotone{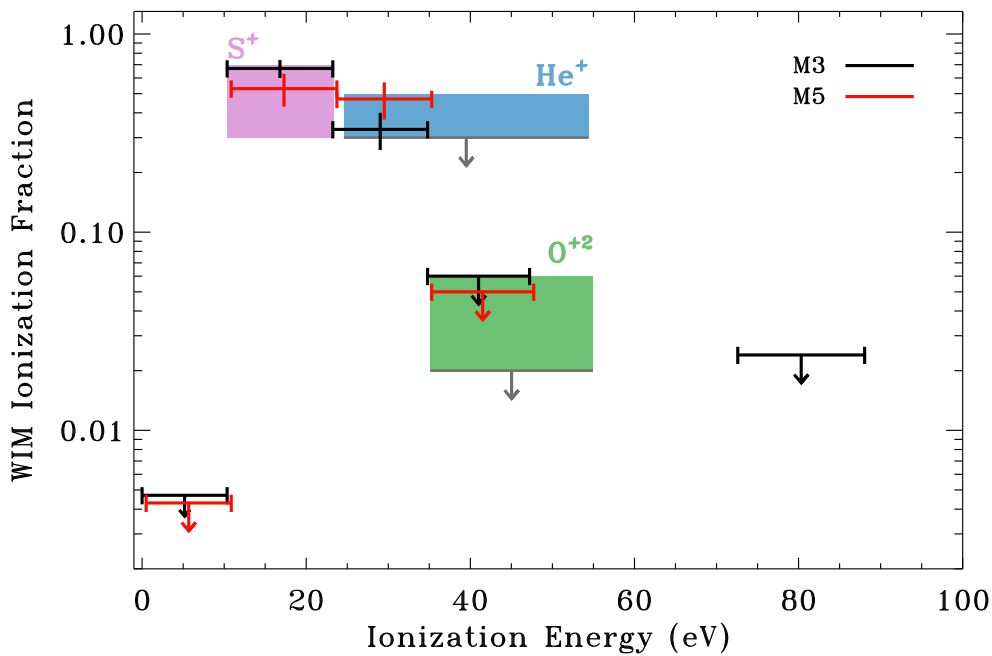}
\caption{A comparison of emission and absorption line constraints on
  ionization fractions for several ions in the WIM.  The ionization
  fractions derived here for S ions are shown for the M~3 (black) and
  M~5 (red) sight lines as in Figure \ref{fig:ionfracs1}.  The shaded
  regions denote the range of values seen by emission line constraints
  on the ionization fractions of of S$^{+}$ (Haffner et al. 1999,
  Madsen et al. 2006 as summarized in Haffner et al. 2009), O$^{+2}$
  (Madsen et al. 2006, Reynolds \& Tufte 1995), and He$^{+}$ (Madsen
  et al. 2006, Madsen 2004, Reynolds 1985b).  We have excluded the
  values associated with the northern filament studied by, e.g.,
  Madsen et al. (2006) that seems to show higher ionization fractions
  for O$^{+2}$ and He$^{+}$ than typically found in the diffuse WIM.
  The lowest upper limits on the emission constraints are shown with
  downward facing arrows. \label{fig:ionfracs2}}
\end{figure}


\begin{figure}
\epsscale{0.9} 
\plotone{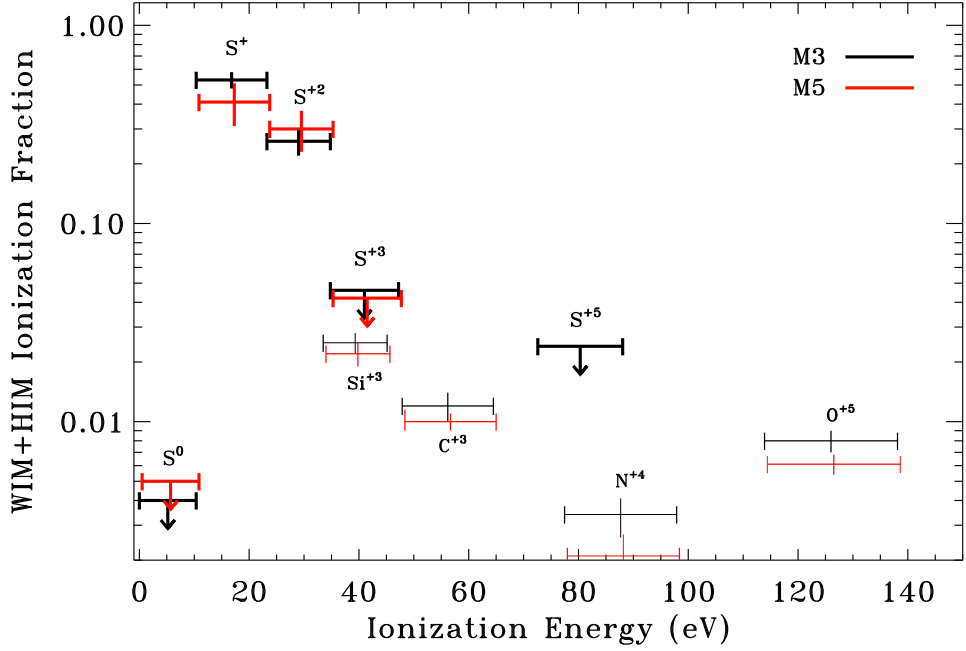}
\caption{The ionization fractions integrated over all of the ionized
  gas (WIM+HIM) of the Milky Way for several metal ions as a function
  of ionization energies for the M~3 (black) and M~5 (red) sight
  lines.  This plot differs from Figure \ref{fig:ionfracs1} in that it
  applies to both the WIM and HIM.  The results are tabulated in Table
  \ref{tab:totalionfracs}.  The mass of ionized gas (at least along
  these sight lines) is dominated by WIM gas traced by ions in the
  range $\sim10$ to 35 eV and by a hotter component with
  $T>4\times10^5$ K probed by ions with higher ionization energies
  than probed by the available UV transitions.  Together the WIM ions
  account for $80\%$ of the total, implying that ions tracing the high
  temperature phase cannot represent more than $\sim20\%$ of the total
  contribution to any metal.  Gas with temperatures in the range a
  few$\, \times10^4$ K to $\sim 4\times10^5$ K contributes very little
  to the total mass budget, as discussed in \S
  \ref{sec:discussion}. \label{fig:totalionfracs1}}
\end{figure}

\begin{figure}
\epsscale{0.9} 
\plotone{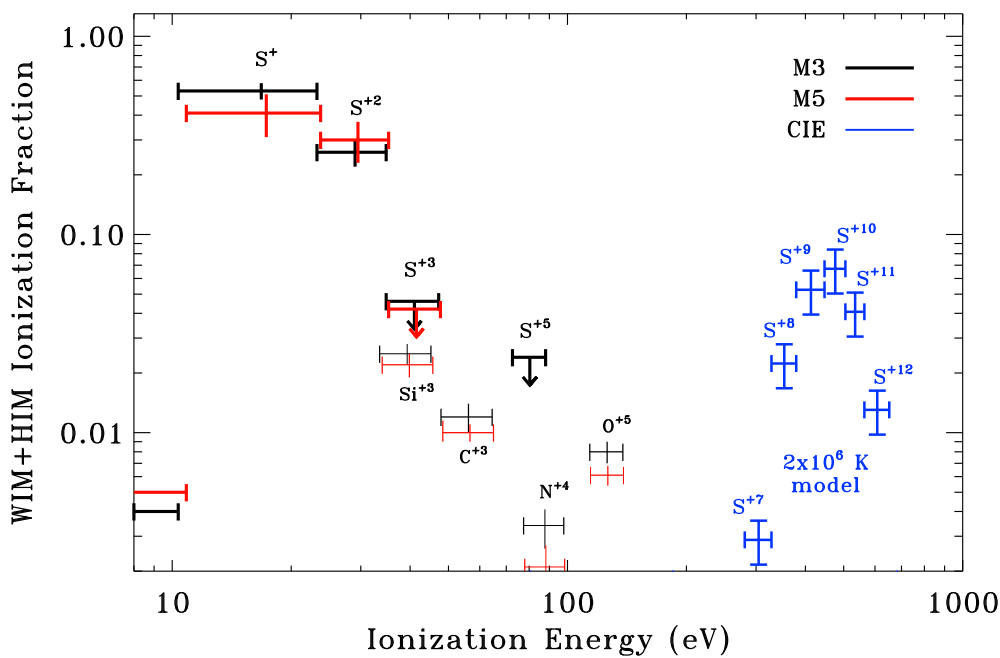}
\caption{The ionization fractions for the total ionized gas (WIM+HIM)
  of the Milky Way, as Figure \ref{fig:totalionfracs1}, for several
  metal ions as a function of ionization energies (plotted on a log
  scale here), including CIE models for S ions in the hot ionized
  phase (blue).  The CIE models assume $T=2\times10^6$ K (e.g., Yao \&
  Wang 2006) for the HIM and have been normalized to contribute
  $(20\pm5)\%$ of the total ionization fraction, although the
  contribution could be somewhat lower.  The ionization fractions of
  all the S ions in this figure sums to unity.  This plot represents a
  map of the energies at which we expect significant ionization
  fraction contributions over all species.  The distribution of S ions
  in CIE will, of course, depend on the assumed
  temperature.  \label{fig:totalionfracs2}}
\end{figure}



\begin{deluxetable}{llccccc}
\tablenum{1}
\tablecolumns{7}
\tablewidth{0pt}
\tablecaption{Adopted Column Densities Toward M~3 and M~5
\label{tab:columns}}
\tablehead{
\colhead{} &
\colhead{} &
\multicolumn{2}{c}{M 3 - vZ 1128} &
\colhead{} &
\multicolumn{2}{c}{M 5 - ZNG 1} \\
\cline{3-4}
\cline{6-7}
\colhead{Species} & 
\colhead{} &
\colhead{$\log N$} &
\colhead{Ref.\tablenotemark{a}} &
\colhead{} &
\colhead{$\log N$} &
\colhead{Ref.\tablenotemark{a}} 
}
\startdata
\ion{H}{1} & & $19.98\pm0.03$ & 1 & & $20.57\pm0.03$ & 4 \\
$e^-$      & & $ 19.91\pm0.01$ & 2 & & $19.96\pm0.01$ & 2,5 \\
%
%
\ion{C}{4} & & $14.39\pm0.03$ & 4 & & $14.38\pm0.02$ & 4 \\ 
\ion{N}{5} & & $13.25^{+0.07}_{-0.09}$ & 4 & & $13.10^{+0.10}_{-0.14}$ & 4 \\
\ion{O}{6} & & $14.49\pm0.03$ & 3 & & $14.41 \pm0.02$ & 4 \\ 
\ion{Si}{4} & & $13.80\pm0.02$ & 4 & & $13.79\pm0.02$ & 4 \\
\ion{S}{1} & & $<12.7 \, (3\sigma)$ & 3 & & $<12.7 \, (3\sigma)$  & 4 \\
\ion{S}{2} & & $15.28\pm0.02$ & 3 & & $15.59\pm0.10$ & 4 \\
\ion{S}{3} & & $14.47\pm0.03$ & 3 & & $14.66\pm0.02$ & 4 \\
\ion{S}{4} & & $<13.7 \, (3\sigma)$ & 3 & & $<13.7 \,(3\sigma)$ & 4 \\
\ion{S}{6} & & $<13.4 \, (3\sigma)$ & 3 & & \nodata & \nodata \\
%
%
%
\enddata
\tablenotetext{a}{References: (1) Howk et al. 2006 ; (2) Hessels et
  al. 2007; (3) Howk et al. 2003; (4) This Work; (5) Freire et
  al. 2008.}

\end{deluxetable}


\begin{deluxetable}{lcc}
\tablenum{2}
\label{tab:results}
\tablecolumns{3}
\tablewidth{0pt}
\tablecaption{Derived Interstellar Parameters Toward 
  M~3 and M~5}
\tablehead{
\colhead{Quantity} &
\colhead{M~3} &
\colhead{M~5} }
\startdata
$\log \Nelectron$      & {$19.91\pm0.01$} & {$19.96\pm0.01$} \\
$\log N(\mbox{\HII})\subhim$ & {$19.20\pm0.18$} & {$18.93\pm0.18$} \\
$\log N(\mbox{\HII})\subwim$  & $19.79\pm0.06$ &  {$19.83\pm0.06$} \\ 
$\log N({\rm H})$\tablenotemark{a} 
                      & $20.24\pm0.03$ &  {$20.66\pm0.02$} \\ 
$N(\mbox{\ion{H}{2}})/N({\rm H})$
                      & $0.45\pm0.09$  & {$0.19\pm0.03$} \\   
$\log A({\rm S})$\tablenotemark{b} 
                      &   $-4.86\pm0.04$ & {$-5.00\pm0.10$}  \\ 
 {[S/H]} & $-0.01\pm0.04$ & $-0.15\pm0.10$ \\
\enddata 
\tablenotetext{a}{The total hydrogen column density, including
  contributions from both warm and hot gas, i.e., including the
  ionized gas associated with the HIM.}
\tablenotetext{b}{The sulfur abundance is derived excluding the \HII\
  column from the HIM.  Thus, the hydrogen reference columns for
  comparison with the summed S ion column densities are $\log N({\rm
    H}) = 20.20\pm0.03$ and $20.64\pm0.02$ for M~3 and M~5,
  respectively.}
\end{deluxetable}


\begin{deluxetable}{lcc}
\tablenum{3}
\label{tab:fractions}
 \tablecolumns{3}
 \tablewidth{0pt}
 \tablecaption{WIM Sulfur Ionization Fractions\tablenotemark{a}}
 \tablehead{
 \colhead{Quantity} &
 \colhead{M~3} &
 \colhead{M~5\tablenotemark{b}} }
 \startdata
 $x({\rm S}^{0})$ & $<0.006 \, (3\sigma)$
                  & $< 0.005 \, (3\sigma)$ \\
 $x({\rm S}^{+})$ & $0.67\pm0.07$\tablenotemark{c} 
                  & $0.53\pm0.09$\tablenotemark{c} \\
 $x({\rm S}^{+2})$ & $0.33\pm0.07$ 
                   & $0.47\pm0.09$ \\
 $x({\rm S}^{+3})$ & $<0.06 \, (3\sigma)$ 
                   & $<0.05 \, (3\sigma)$ \\
 $x({\rm S}^{+5})$ & $<0.03 \, (3\sigma)$
                   & \nodata \\
\enddata
\tablenotetext{a}{The ionization fractions are $x(X^j)\equiv
  N(X^j)/N(X)$.  We assume $x({\rm H}^+)=0.95\pm0.05$ in deriving
  these values (see text).}
\tablenotetext{b}{We adopt the abundance towards M~3 for the
  ionization fraction calculations.  If one adopts the less
  well-determined value derived for M~5, the result for S$^{+2}$, for
  example, is $x({\rm S}^{+2})=0.67\pm0.19$.}
\tablenotetext{c}{The S$^+$ ion fractions are not measured directly;
  instead they are derived assuming $x({\rm S}^{+}) + x({\rm S}^{+2})
  = 1$ (see text).} 
 \end{deluxetable}

\begin{deluxetable}{lcc}
\tablenum{4}
\label{tab:totalionfracs}
 \tablecolumns{3}
 \tablewidth{0pt}
 \tablecaption{Total (WIM+HIM) Ionization Fractions\tablenotemark{a}}
 \tablehead{
 \colhead{Quantity} &
 \colhead{M~3} &
 \colhead{M~5} }
 \startdata
 $x({\rm S}^{0})$ & $<0.005 \, (3\sigma)$\tablenotemark{c} 
                  & $< 0.004 \, (3\sigma)$\tablenotemark{c} \\
 $x({\rm S}^{+})$ & $0.53\pm0.10$\tablenotemark{d} 
                  & $0.42\pm0.10$\tablenotemark{d} \\
 $x({\rm S}^{+2})$ & $0.27\pm0.04$\tablenotemark{c} 
                   & $0.38\pm0.08$\tablenotemark{c} \\
 $x({\rm S}^{+3})$ & $<0.05 \, (3\sigma)$\tablenotemark{c} 
                   & $<0.04 \, (3\sigma)$\tablenotemark{c} \\
 $x({\rm S}^{+5})$ & $<0.02 \, (3\sigma)$\tablenotemark{d} 
                   & \nodata \\
 $x({\rm C}^{+3})$ & $0.012^{+0.002}_{-0.0008}$\tablenotemark{d} 
                   & $0.10^{+0.001}_{-0.0007}$\tablenotemark{d} \\     
 $x({\rm N}^{+4})$ & $0.003\pm0.001$\tablenotemark{e} 
                   & $0.002\pm0.002$\tablenotemark{e} \\
 $x({\rm O}^{+5})$ & $0.008\pm0.001$\tablenotemark{e} 
                   & $0.006\pm0.001$\tablenotemark{e} \\
 $x({\rm Si}^{+3})$ & $0.025\pm0.03$\tablenotemark{e}
                    & $0.022\pm0.003$\tablenotemark{e} \\
             
\enddata

\tablenotetext{a}{The ionization fractions are $x(X^j)\equiv
  N(X^j)/N(X)$.  We assume $x({\rm H}^+)=0.95\pm0.05$ (see text).
  For the hottest gas, we expect this fraction to be unity.}
\tablenotetext{b}{We adopt the abundance towards M~3 for the
  ionization fraction calculations.}
\tablenotetext{d}{The S$^+$ ion fractions are not measured directly;
  instead they are derived assuming $x({\rm S}^{+}) + x({\rm S}^{+2})
  = 1$ (see text).} 
\tablenotetext{e}{Ions higher than ${\rm S}^{+3}$ have an assumed
$x({\rm H}^+)=1.0$}

 \end{deluxetable}

\end{document}